\documentstyle[12pt]{article}

\begin{document}
\begin{titlepage}

\begin{flushright}
UAB-FT-322\\
September 1993
\end{flushright}

\vspace{\fill}

\begin{center}
        {\LARGE\bf  LOOP ACTION FOR LATTICE U(1) GAUGE THEORY}
\end{center}

\vspace{\fill}

\begin{center}
       { {\large\bf
        J. M. Aroca }
	\vskip 0.5cm
        Departament de Matem\`atiques, \\
        Universitat Polit\`ecnica de Catalunya, \\
        Escola T\`ecnica Superior d'Enginyers de Telecomunicaci\'o, E-08034 \\
        Barcelona, Spain
        \ \\
\vspace{ 7 mm}
        and
\vspace{ 7 mm}
        \ \\
        {\large\bf  H. Fort}
	\vskip 0.5cm
        Grup de F\'\i sica Te\`orica\\
        and\\
        Institut de F\'\i sica d'Altes Energies\\
        Universitat Aut\`onoma de Barcelona\\
        08193 Bellaterra (Barcelona) Spain}
\end{center}

\vspace{\fill}

\begin{abstract}
It is showed that the very recently introduced Lagrangian $loop$ formulation
of the lattice Maxwell theory is equivalent to the Villain form
in 2+1 dimensions. A
transparent description of the classical $loop$ action is
given in pure geometrical terms for the $2+1$ and $3+1$ dimensional cases.
\end{abstract}

\end{titlepage}

The $loop$ formalism was introduced some time ago as a
general analytical Hamiltonian approach based on the properties
of the $group$ $of$ $loops$ \cite{gt}.
It provides a common framework for quantum gauge theories
-- it works for several models, as Maxwell theory \cite{gt1},
Chern-Simons theory\cite {l}, etc --
and quantum gravity \cite{ahrss}.
Furthermore,
it works for Yang-Mills theories on a lattice \cite{gt2} and
recently has been extended in such a way to take account of
dynamical fermions \cite{fg}.
Very recently \cite{aggs} a loop action for the U(1) gauge
theory   has been built which
lattice version leads to the Kogut-Susskind Hamiltonian.
A Lagrangian formulation in terms of loops
is interesting for multiple reasons. First, it offers
the possibility of knit together the intrinsic advantages of
the nonlocal loop description (the non-redundancy
of gauge degrees of freedom and its geometrical transparency) and
the computational power of numerical simulations. This provides a
a useful complement to the analytical loop studies.
The classical action may be also relevant to perform semiclassical
approximations.
In this paper we show that the proposed action for the D=2+1
dimensional case
leads after a duality transformation to the
discrete Gaussian model form. It is known that the same happens
for the Villain form. Thus, for D=2+1, the minimal description
provided by the loops actually corresponds to that of Villain form
instead of Wilson cosinus form. For D=3+1 the $loop$ action and Villain
form give a similar description but the connection between them
is more subtle.

According to reference \cite {aggs}
the continuum Euclidean loop action for the abelian theory in $D=d+1$
space-time dimensions is given by
\begin{equation}
 S=\frac{g^2}{2} \int dtd^dx \{-\dot{X}_C^i({\bf x})
        \frac{1}{\Delta}\dot{X}_C^i({\bf x})
         + X_C^i({\bf x})X_C^i({\bf x})\}
\label{1}
\end{equation}
where $X_C^i({\bf x})$ is the loop current
$\oint _Cdy^i\delta^3({\bf x}-{\bf y})$ and $\Delta$ is the
three-dimensional Laplace operator. The time derivative of
$X_C^i({\bf x})$ can be written as
$\lim_{dt\rightarrow 0}\\
\frac{1}{dt}\oint _{C_{t+dt}
\bar{C}_t}dy^i\delta^3({\bf x}-{\bf y})$, where
$\bar{C}_t$ denotes the loop $C_t$ traversed in the opposite
direction.

In order to formulate (\ref{1}) on a lattice we represent the
continuum
surface spanned by the loop $C$ as a set of spatial loops
$C_t$ at different times  $t=0,1,\ldots ,T$. We also replace the
derivatives by finite difference operators and get

\begin{equation}
 S=\frac{g^2}{2} \sum_t
 \sum_s\{-X_{C_{t+1}\bar{C}_t}^i(s)
 \frac{1}{\Delta}X_{C_{t+1}\bar{C}_t}^i(s)
 + X_{C_t}^i(s)X_{C_t}^i(s)\}
\label{2}
\end{equation}
where $X_{C_t}^i(s)$ is an integer which counts the number of times
that the loop $C_t$ traverses the link $(s,\hat{i})$.

We shall consider first the D=$2+1$ dimensional case. The loops
$C_t$ lie on a plane and have a unique decomposition in
plaquettes $C_t=\prod_p p^{n_p}$, where $n_p$ is the plaquette
multiplicity, i.e. the number of times that plaquette $p$ is
contained in loop $C$. The first term in (\ref{2}) is then

\begin{equation}
\sum_s-X_D^i(s) \frac{1}{\Delta}X_D^i(s)=
\sum_{p,p'}\sum_s-n_pn_{p'}X_p^i(s) \frac{1}{\Delta}X_{p'}^i(s)=
\sum_{p,p'}n_pn_{p'}l(p,p')
\end{equation}

where $D$ denotes the loop $C_{t+1}\bar{C}_t$, $l(p,p')$ is the
interaction between two plaquettes and is given by

\begin{equation}
l(p,p')\equiv \sum_s-X_p^i(s) \frac{1}{\Delta}X_{p'}^i(s)=
\sum_{s,s'}X_p^i(s)G(s,s')X_{p'}^i(s')
\end{equation}

with

\begin{equation}
G(s,s')= \int_0^{2\pi } \frac{d^2q}{(2\pi)^2}
\frac{e^{iq(s-s')}}{2(2-\cos q_1-\cos q_2)}
= \delta_{s,s'}.
\end{equation}

That gives

\begin{equation}
\sum_s-X_D^i(s) \frac{1}{\Delta}X_D^i(s)=
        \sum_{p,p'}n_pn_{p'}\delta_{p,p'}=
        \sum_p n_p^2=
        A_2(D)
\end{equation}
where $A_2(D)$ actually denotes the quadratic area, i.e. the sum of
the squares of the constitutive pla\-que\-tte mul\-ti\-pli\-ci\-ties,
of the surface $\cal S$ enclosed by the loop $D$.

The second term on the action is

\begin{equation}
\sum_sX_C^i(s)X_C^i(s)=\Lambda(C)
\end{equation}
where $\Lambda(C)$ denotes the quadratic length --the sum of
squares of link mul\-ti\-pli\-ci\-ties-- of the loop $C$.

We arrive then to the following geometric expression for
the lattice action:

\begin{equation}
 S=\frac{g^2}{2}\sum_t
\{ A_2(C_{t+1}\bar{C}_t) + \Lambda(C_t) \}.
\label{Loop3}
\end{equation}

Now, instead of expressing the loops as an integer field defined
on the links of the lattice, we express them through the surface
they enclose by giving an integer field defined on the spatial
plaquettes.
By duality this field is defined on the sites $(\tilde{x},\tilde{y},t)$
dual
to these spatial plaquettes so the configurations are given by integer
fields $k(\tilde{x},\tilde{y},t)$. The geometric terms in the action
are expressed then as

\begin{equation}
\Lambda (C_t)=\sum_{\tilde{x},\tilde{y},t}
\{(k(\tilde{x}+1,\tilde{y},t)-k(\tilde{x},\tilde{y},t))^2+
        (k(\tilde{x},\tilde{y}+1,t)-k(\tilde{x},\tilde{y},t))^2\}
\end{equation}

and

\begin{equation}
A_2(C_{t+1}\bar{C}_t)=\sum_{\tilde{x},\tilde{y},t} (k(\tilde{x},\tilde{y},t+1)
-k(\tilde{x},\tilde{y},t))^2.
\end{equation}

Collecting these results, the action reads finally

\begin{equation}
S=\frac{g^2}{2} \sum_{s,\mu} (k(s+\hat{\mu})-k(s))^2
\label{12}
\end{equation}
which is the action of the integer Gaussian spin model.
In this expression the covariance hidden in (\ref{2}) is manifest.

On the other hand, the duality of the Villain form
of the U(1) gauge theory to discrete Gaussian
models --discrete Gaussian spin model for D=3, discrete Gaussian gauge
model for D=4-- was already established \cite{bkm}, \cite{S}.
Thus, for D=3, we have found that the minimal description that provides the
loop
action (2) is equivalent to the Villain form since that
both lead by duality to (12).


With the aim to explore this connection between loops and Villain form
for an arbitrary dimension D we shall resort to the language of
forms. Working in this general framework we will show that we recover
the previous equivalence for D=2+1 and we will analyze in detail the
relation between both theories at D=3+1.

A p-form is a function defined on the p-cells of the lattice
(p=0 sites, p=1 links, p=2 cubes, etc.) over an abelian group
which shall be {\bf R}, {\bf Z}, or U(1)={reals module 2$\pi$}.
Integer forms can be considered geometrical objects on the lattice.
For instance, a 1-form is a path and the integer value on a link
is the number of times that the path traverses this link.
$\nabla$ is the co-border operator which
maps p-forms onto (p+1)-forms. It is the gradient operator when acting
on scalar functions (0-forms) and it is the rotational on vector functions
(1-forms). We shall consider the scalar product of p-forms defined
$<\alpha \mid \beta> = \sum_{c_p}\alpha (c)\beta (c)$ where the sum runs
over the p-cells of the lattice. Under this product the $\nabla$
operator is adjoint to the border operator $\partial$ which maps
p-forms onto (p-1)-forms and which corresponds to minus times the usual
divergence operator. The operator $\Box =\nabla \partial +\partial \nabla$
is called the Laplacian and differs only by a minus sign of the current
Laplacian $ \Delta_\mu \Delta_\mu$.
We shall use also the duality transformation which maps biyectively p-forms
over (D-p)-forms. Under duality the border and co-border operators
interchange.

In the loop theory exposed before, physical configurations correspond
to distributions of spatial loops at different times. These are
spatial closed integer 1-forms. Then we consider only forms $c$ which
are 0 for temporal links and closed ($\partial c=0$).
The border and co-border operators can be restricted to the spatial sections
(t=constant) and we denote the spatial Laplacian as $\Delta$.
The loop action (2) can be expressed as

\begin{equation}
S=\frac{g^2}{2}\sum_t  (<c_{t+1}\bar{c}_t,\frac{1}{\Delta}c_{t+1}\bar{c}_t>+
               <c_t,c_t>)=
\frac{g^2}{2} <c,\frac{\Box}{\Delta}c>
\label{}
\end{equation}

and the path integral

\begin{equation}
Z =   \sum_{
                \begin{array} {c} c\\
                              \left(  \partial c = 0 \right)
                \end{array}
                 }\exp
(-\frac{g^2}{2} <c,\frac{\Box}{\Delta}c>)
\end{equation}
where c denotes a spatial loop i.e. c belongs to a t=constant slice.

It is possible to solve the constraint $\partial c=0$
by taking $c=\partial s$ where s are integer spatial 2-forms.

\begin{equation}
Z =   \sum_{s}{'}\exp
(-\frac{g^2}{2} <\partial s,\frac{\Box}{\Delta}\partial s>)
\end{equation}

The prime on the sum indicates that we must restrict the $s$'s
over which we sum since we have introduced the symmetry
$s \rightarrow s+\partial
v$ for arbitrary 3-forms $v$ (in other words, there are
infinite surfaces with the same loop as border).
This restriction is the usual gauge-fixing problem and is
solved more simply if we make a duality transformation
in the spatial sections. In terms of the spatial (D-3)-forms $*s$
(14) is written as

\begin{equation}
Z =   \sum_{*s}\exp
(-\frac{g^2}{2} <\nabla *s,\frac{\Box}{\Delta}\nabla *s>)
\end{equation}

For D=2+1 the dual of $*s$ are 0-forms $*s$.
We do not need to fix the gauge since a spatial loop in the
border of a unique surface.

If we use the Poisson sumation formula
$\sum_j f(j) = \sum_k \int_{-\infty}^{\infty} d\phi f(\phi) e^{2\pi i\phi k}$
we can substitute the integer $*s$ variables
by integer 0-forms $*m$ and real 0-forms $\phi$, and performing
the Gaussian integration over $\phi$ we get

\begin{equation}
Z =   \sum_{m} \exp
(-\frac{2\pi^2}{g^2} <*m,\frac{1}{\Box}*m>)
\end{equation}

The above expression is a sum over point monopoles which interact
through a Coulomb potential. Its form is equal to the one obtained
from the Villain form of the U(1) theory.
Those monopoles are defined in the centers of the spatial plaquettes
and these configurations
are in correspondence one-to-one with scalar forms on the
three-dimensional lattice. In this case, as we have seen before,
the loop theory describes
exactly the topological excitation of the Villain theory.

\vspace{5 mm}

For the D=3+1 case the dual of the $s$'s in (14) are spatial
1-forms $*s$.
To fix $*s$ we impose $*s(l)=0$ for every link $l\in T$,
$T$ being a maximal
spatial tree. This selects only one $*s$ among all $*s$'s with the
same value of $\nabla *s$.
If we use the Poisson sumation formula
we can substitute the integer $*s$ variables on the complementary of $T$,
$\bar{T}$ by integer 1-forms $*m$ and real 1-forms $\phi$.
Once again the $\phi$ integration gives




\begin{equation}
Z = constant \sum_{
                \begin{array} {c} *m\\
                              \left(  \partial *m = 0 \right)
                \end{array}
                 }\exp
(-\frac{2\pi^2}{g^2} <*m,\frac{1}{\Box}*m>)
\end{equation}

i.e. a sum over {\em spatial} closed loops which interact
through a Coulomb potential.
The Villain form of the U(1) gauge theory leads to the same
expression but the monopole loops
are not restricted to lie at t=constant.
 Thus, the loop
theory is not fully equivalent to the Villain U(1) theory at D=4.
\vspace {3 mm}


        We can deepen our insight on the relation between the
$loop$ and the Villain theory following the opposite way: starting
from Villain
form and then repeating the previous treatment in terms of p-forms.
The path integral for the Villain U(1) theory is

\begin{equation}
Z = \int (d\theta ) \sum_{ j }\exp
(-\frac{\beta}{2}\mid\mid \nabla \theta -2\pi j\mid\mid^2)
\label{}
\end{equation}

where $\theta$ is a real periodic 1-form, that is,
a real number $\theta_l \in [0, 2\pi ]$
defined in each link of the lattice. $j$ are integer 2-forms, defined
at the lattice plaquettes, and $\mid\mid . \mid\mid^2 = <.,.>$.
Using the Poisson sumation formula
and integrating first in the continuum $\phi$ variables and then
over the compact $\theta$, we finally get



\begin{equation}
Z = (2\pi \beta)^{-N_p/2}  \sum_{
                \begin{array} {c} s\\
                              \left(  \partial s = 0 \right)
                \end{array}
                 }\exp
(-\frac{1}{2\beta}<s,s>)
\propto \sum_{s}\exp (-\frac{1}{2\beta}A_2({\cal S})).
\label{A2}
\end{equation}

where $N_p$ in the number of plaquettes of the lattice.
We can now analyze the geometrical meaning of this equation.
$s$ are integer 2-forms, that is they are defined over
bidimensional surfaces on the
lattice. The condition $\partial s=0$ means to sum only over
closed surfaces, the action $<s,s>$ is proportional to the
quadratic area of the surface.
Selecting one direction of the lattice as a `time' direction
one can express $<s,s>$ as (see Appendix)

\begin{equation}
<s,s>= \sum_t  (<c_t,c_t> +
   <c_{t+1}\bar{c}_t,\frac{1}{\Delta}c_{t+1}\bar{c}_t>+
   <\nabla {\bf s}_t,\frac{1}{\Delta}\nabla {\bf s}_t> )
\end{equation}

where ${\bf s}_t$ represents a pure spatial 2-form.
The first two terms in (20) correspond exactly
to the loop action (12). If we work at D=2+1 we have only one solution
to $\partial {\bf s}_t=c_{t+1}\bar{c}_t$. In this case the
spatial co-border of a 2-form is 0, then the third term in (20)
vanishes and we obtain the loop theory (12).

For D=3+1 the loop action can not take account of the term
$<\nabla {\bf s}_t,\frac{1}{\Box}\nabla {\bf s}_t>$.
The Lagrangian lattice loop description seems to miss
some degrees of freedom present in the gauge
U(1) theory (the general monopole loops which are not contained
in the t=constant hypersurfaces).
As it was previously mentioned it has been shown that
the loop action leads to the Kogut-Susskind Hamiltonian so for
continuous time both descriptions are equivalent.


        It seems interesting to have  a Lagrangian lattice loop
description to continue exploring some interesting points. For example,
the pure 3+1 U(1) gauge theory  was studied by
means of Hamiltonian analytical method in
the loop representations \cite{af1}. In this reference we found
a continuous phase transition for this model instead of the
first order measured by means of standard Monte Carlo simulations.
A possible explanation for this difference relies on the fact that
we have used a cluster method which implies a formally infinite
lattice which is free from the spurious topological effects occurring
in finite lattices with periodic boundary conditions \cite{gr} \cite{gup}.
This periodicity leads, on the confining side, to large monopole loops
which wrap around the lattice. Those loops clearly have a fairly large
action associated with them, so a jump is observed when they disappear
at deconfining transition. This jump no longer survives in the infinite
volume limit. The effect of this wrapping loops in driving the
transition is crucial.
Recently, Stack and Wensley \cite{sw} have calculated the heavy quark
potential from the magnetic current due the monopoles. They have resolved
the magnetic current into large loops which wrap around the lattice
and simple loops which do not. They found that the long range part of the
heavy quark potential can be calculated solely from the large, wrapping
loops of magnetic current.

 The loop's method seems to be in a good position because the
finite lattice is not one of its intrinsic features.
The development of different Lagrangian computational techniques
for studying the D=4 U(1) model in order to complement our previous
Hamiltonian results is one of our present aims. Also, we are
interested in extending the loop's action in such a way to include
matter fields.

\vspace {0.2cm}

We wish to thank, D.Armand-Ugon, R.Gambini, M.Grady and L.Setaro for
useful discussions and comments.

\vspace {0.5cm}

\appendix{\bf APPENDIX}

	Selecting a direction on the lattice which we call time
and consider the division of the lattice in space sections of
dimension D-1. Any surface 2-form $s$ can be written
$s=\sum_t({\bf s}_t+\tilde{s}_t)$ where ${\bf s}$ is purely spatial and
$\tilde{s}$ is temporal (its plaquettes have one link in the $t$
direction joining the space sections $t$ and $t+1$). The $t$
subindex correspond to each spatial slice
of the lattice. If we consider the borders of these surfaces we
have that they are always spatial 1-forms (loops) in the
slice $t$ since the temporal parts of $\partial \tilde{s}_t$ cancel.
$\partial \tilde{s}_t$  has two
spatial components in the sections $t$ and $t+1$,
$(\partial \tilde{s}_t)_{inf}$ and $(\partial \tilde{s}_t)_{sup}$.
We define $c_t=(\partial \tilde{s}_t)_{sup}$. Then
$(\partial \tilde{s}_t)_{inf} = -c_t$.
Imposing $s$ to be closed we have $\partial {\bf s}_t=c_{t+1}-c_t$.
We note also that if $p$ is a plaquette defined by a
spatial link $l$ and a temporal link, then the integers
$\tilde{s}_t(p)$ and $c_t(l)$ coincide.
The quadratic area is then
$$ <s,s>=\sum_t  (<\tilde{s}_t,\tilde{s}_t> + <{\bf s}_t,{\bf s}_t> )= $$
$$ \sum_t  (<\tilde{s}_t,\tilde{s}_t> +
   <\partial {\bf s}_t,\frac{1}{\Delta}\partial {\bf s}_t>+
   <\nabla {\bf s}_t,\frac{1}{\Delta}\nabla {\bf s}_t> ) = $$
\begin{eqnarray}
\sum_t  (<c_t,c_t> +
   <c_{t+1}\bar{c}_t,\frac{1}{\Delta}c_{t+1}\bar{c}_t>+
   <\nabla {\bf s}_t,\frac{1}{\Delta}\nabla {\bf s}_t> )
\end{eqnarray}

In the second equality we have inserted in the second term
$\Box \Box^{-1}$ and used the adjoint relation between
border and coborder operators. In this expression the
operators belong to the spatial hypersurface, so they 'live'
in dimension D-1.
In the path integral the sum over $s$ is replaced for
a sum over the distributions of spatial loops $c_t$ and
spatial surfaces $s_t$ such that $\partial {\bf s}_t=c_{t+1}\bar{c}_t$.

\end{document}